\documentclass[prx,aps,onecolumn,10pt]{revtex4-1}

\usepackage{booktabs,amsmath,amssymb,graphicx,xcolor,esint,multirow}

\newcommand{\Ei}{{\bf E}_{i}}
\newcommand{\Esd}{\tilde{\bf E}_{S}^{+}}
\newcommand{\Est}{\tilde{\bf E}_{S}^{-}}

\newcommand{\re}[1]{\mbox{Re}\left\{#1\right\}}

\newcommand{\no}{\hat{\bf n}}

\newcommand{\J}[2]{j_{#2}\left(#1\right)}
\newcommand{\Etoto}{{\bf E}_0}

\newcommand{\SqrtEps}{\sqrt{\varepsilon_r}}

\newcommand{\Mi}[3]{{\bf M}_{#3} ^{\left(#1\right)} \left( #2 \right)}
\newcommand{\Ni}[3]{{\bf N}_{#3}^{\left(#1\right)} \left( #2 \right)}
\newcommand{\h}[2]{h_{#2}^{\left( 1 \right)}\left(#1\right)}

\begin{document}

\title{Directional Scattering Cancellation for an Electrically Large Dielectric Sphere}

\author{Carlo Forestiere}
\affiliation{ Department of Electrical Engineering and Information Technology, Universit\`{a} degli Studi di Napoli Federico II, via Claudio 21,
 Napoli, 80125, Italy}
\author{Giovanni Miano}
\affiliation{ Department of Electrical Engineering and Information Technology, Universit\`{a} degli Studi di Napoli Federico II, via Claudio 21,
 Napoli, 80125, Italy}
\author{Mariano Pascale}
\affiliation{ Department of Electrical Engineering and Information Technology, Universit\`{a} degli Studi di Napoli Federico II, via Claudio 21,
 Napoli, 80125, Italy}
\author{Roberto Tricarico}
\affiliation{ Department of Electrical Engineering and Information Technology,
	 Universit\`{a} degli Studi di Napoli Federico II, via Claudio 21,
 Napoli, 80125, Italy}

\begin{abstract}

We demonstrate the directional scattering cancellation for a dielectric sphere of radius up to ten times the incident wavelength, by coating it with a surface of finite conductivity. Specifically, the problem of determining the values of the surface conductivity that guarantees destructive interference among hundreds of multipolar scattering orders at the prescribed angular direction is reduced to the determination of the zeros of a polynomial, whose coefficients are analytically known . 
 \end{abstract}

\maketitle

More than three decades ago, Kerker and co-workers predicted  the suppression of the directional scattering in magneto-dielectric spheres of arbitrary size, illuminated by a plane wave.
Specifically, they theoretically demonstrated backscattering cancellation if $\varepsilon = \mu$, and nearly zero forward scattering if $\varepsilon = \left(4 - \mu \right)/\left(2\mu +1 \right) $ \cite{kerker1983electromagnetic}. In both conditions, the scattering cancellation results from the destructive interference between magnetic and electric multipoles of same order. More recently, the backscattering cancellation in a {\it small} non-magnetic sphere  was demonstrated  \cite{gomez2011electric,nieto2011angle}  by engineering the interference of magnetic and electric dipoles. This scenario, that generalizes the first Kerker's condition, has been experimentally observed both in the microwaves \cite{geffrin2012magnetic} and in the visible spectral range \cite{fu2013directional,Person13}. 
Rigorous conditions for the directional scattering cancellation may have a great impact in scattering shaping and control, in optical wireless nano-antenna links, in manipulation of quantum dot emission, and in the engineering of optical polarizations and forces.

However, as stated in a 2018 review on generalized Kerker effects in nanophotonics and metaoptics \cite{liu2018generalized}, ``almost all previous studies have only explored interference of low-order multipoles (up to quadrupole) whereas high-order multipole can certainly bring extra opportunities''. The main reason behind this fact is that directional interference can be analytically designed only when a small number of multipoles is involved, while, for larger objects, it has been necessary so far to resort to trial-and-error or optimization algorithms.
In \cite{Forestiere16,pascale2017spectral} there is a first attempt to devise a general method to cancel the backscattering of a homogeneous sphere that applies beyond the small particle regime. However, the cancellation of the backscattering was only limited to a sphere of radius of the order of the incident wavelength, due to a lack of robustness that appears when many scattering orders are included.

Here, by using a surface coating, we are able to  cancel the directional scattering of a non-magnetic sphere of radius up to ten times the incident wavelength. The surface coating may be implemented by a two-dimensional material \cite{xia2014two}, or by a film with thickness much smaller than the sphere radius and than the wavelength .
The method we used leverages on an analytical decomposition of the full wave scattered field in terms of material independent modes \cite{Forestiere16}. In this case, both poles and modes are found analytically. This fact allows to overcome the limitations of Refs. \cite{Forestiere16,pascale2017spectral}. Then, we determine the values of the surface conductivity that guarantee the destructive interference among hundreds of multipolar scattering order, by finding the zeros of a polynomial, whose coefficients are analytically known. This approach also enables the identification of the modes responsible for the interference.
We demonstrate the validity of the introduced approach by cancelling the backscattering of a dielectric sphere of radius varying from one-fourth to ten times the incident wavelength, and the forward scattering of a sphere of radius varying from one-fourth to five times the incident wavelength.

Let us consider the electromagnetic scattering by a sphere of radius $r$, occupying the region $\Omega^+$, of boundary $\Sigma= \partial \Omega^+$.
In the following we denote the external region with $\Omega^-$. The object is excited by a time harmonic electromagnetic field incoming from infinity, namely $\re{\Ei \left({\bf r}\right) e^{- i \omega t}}$, where $\omega=2\pi c_0/\lambda$ is the frequency, $\lambda$ is the wavelength, and $c_0$ is the speed of light in vacuum.  The material of the object is a non-magnetic, isotropic, homogeneous dielectric with relative permittivity $\varepsilon_{r}$, surrounded by vacuum. We denote with ${\bf E}_{0}^\pm$ the total field in $\Omega^\pm$, and the corresponding scattered field as:
\begin{equation}
{\bf E}_{S, \,0}^\pm = 
   {\bf E}_{0}^\pm - {\bf E}_{i}.
\end{equation}
If the incident field is a x-polarized plane wave, propagating along $z$, i.e. ${\bf E}_i =  {E}_{inc} \hat{\bf x} \, e^{ikz}$, where $k = \omega/c$, the scattered field in $\Omega^-$ can be expressed in terms of vector spherical wave functions (VSWFs) \cite{bohren08} as
\begin{equation}
   {\bf E}_{S,0}^- \left( {\bf r} \right)  =  \displaystyle\sum_{n=1}^\infty  E_n \left(  i a_n^0 \Ni{3}{k_0 {\bf r}}{e1n} -b_n^0 \Mi{3}{k_0 {\bf r}}{o1n}   \right),
  \label{eq:PW_VSWF}
\end{equation}
where $E_n= i^n E_{inc} \left( 2 n + 1 \right)/n \left( n + 1 \right)$, the subscripts $e$ and $o$ denote even and odd, while the expression for $a_n^0$ and $b_n^0$ can be found in \cite{bohren08}.

Now, in order to modify the scattering properties of this object, we cover the domain $\Omega^+$ with a surface coating. The coating is made of a linear, homogeneous, isotropic, time-dispersive material of finite conductivity $\sigma \left( \omega \right) $.  We denote with ${\bf E}^\pm$ the total field in this modified scenario. The constitutive relation of the 2D coating material is
\begin{equation}
{\bf J}_s = \sigma \overleftrightarrow{\mbox{T}} {\bf E}^-,
\label{eq:ConstRel}
\end{equation}
where ${\bf J}_s$ is the surface current density on $\Sigma$, and $\overleftrightarrow{\mbox{T}}$ is the projector that extracts the tangential component of the electric field to the oriented surface $\Sigma$, namely $
   \overleftrightarrow{\mbox{T}} \left( \cdot \right) = - {\bf n} \times \left[ {\bf n} \times \left( \cdot \right) \right]   
$.
The scattered field in $\Omega^\pm$, denoted as ${\bf E}_S^{\pm}$ , is defined as:
$
{\bf E}_{S}^{\pm} = 
   {\bf E}^\pm - {\bf E}_{i}.
$
It can be decomposed as 
$
 {\bf E}_S^{\pm} = \tilde{\bf E}_{S}^{\pm} +  {\bf E}_{S,0}.
$
The field $\tilde{\bf E}_{S}^{\pm}$ represents the change in the electric field caused by the introduction of the coating, namely
\begin{equation}
 \tilde{\bf E}_{S}^{ \pm }   =  {\bf E}_S^{ \pm } - {\bf E}_{S,0}^\pm = {\bf E}^\pm - {\bf E}_{0}^\pm.
\end{equation}
The field  $ \tilde{\bf E}_{S}^{\pm}$ is solution of the following problem:
\begin{alignat}{2}
\label{eq:MErot2}
& k_0^{-2} \boldsymbol{\nabla}^2 \Esd + \varepsilon_{r} \left( \omega \right) \Esd =  {\bf 0}  \quad && \mbox{in} \, \Omega^{+}, \\
\label{eq:MErot3}
& k_0^{-2} \boldsymbol{\nabla}^2 \Est + \Est = {\bf 0}  \; && \mbox{in} \, \Omega^{-}, 
\end{alignat}

\begin{equation}
   \begin{aligned}
   \label{eq:BC2}
   & \no \times \left( \Est - \Esd \right) = {\bf 0} \\
   & \no \times \left( \boldsymbol{\nabla} \times \Est - \boldsymbol{\nabla} \times\Esd \right) = i \omega \mu_0 \sigma \overleftrightarrow{\mbox{T}} \, \left( {\bf E}_{0}^- + \Est \right) \\
\end{aligned} \qquad   \mbox{on} \, \Sigma. \\
\end{equation}
Equations \ref{eq:MErot2}-\ref{eq:BC2} have to be solved with the radiation conditions, which constraint the scattered field to be an outgoing wave.

Let us consider the auxiliary homogeneous problem obtained from Eqs. \ref{eq:MErot2}-\ref{eq:BC2} by zeroing the driving term ${\bf E}_{0}^-$. 
The electric field, that satisfies the homogeneous problem, can be represented in terms of VSWFs  in the domains $\Omega^+$ as:
\begin{equation}
 \displaystyle\sum_{n=1}^\infty \sum_{m=0}^n \sum_{p\in \left\{e,o\right\}} \left\{ C_{pmn} \Mi{1}{ \SqrtEps k_0  {\bf r}}{pmn} +  D_{pmn} \Ni{1}{ \SqrtEps k_0  {\bf r}}{pmn} \right\} 
 \label{eq:Cp} 
\end{equation}
and in  $\Omega^+$ as:
\begin{equation}
    \displaystyle\sum_{n=1}^\infty\sum_{m=0}^n \sum_{p\in \left\{ e, o\right\}} \left\{ A_{pmn} \Ni{3	}{k_0 {\bf r}}{pmn} + B_{pmn} \Mi{3}{k_0 {\bf r}}{emn} \right\}.
\label{eq:Cm} 
\end{equation}

\begin{table*}
\centering
\begin{tabular}{cccccccc} 
  \hline
 \multirow{2}{*}{$r/\lambda$} &     \multirow{2}{*}{$N_{max}$} &  \multirow{2}{*}{$\zeta_0 \sigma$} & 
   \multicolumn{5}{c}{$ \left\| \mathbf{E} \left( \theta = 0 \right)/\mathbf{E}\left( \theta = \pi \right) \right\| (dB) $}    \\
 \cmidrule(l){4-8}
   & & & $sf=1$ & $sf=2$&$sf=3$ &    $sf=4$& $sf=5$  \\
   \hline
 0.25 & $10$ & $0.19539 + 0.43197 \, i$ &  $35$& $52$ & $74$ & $95$ & $113$  \\
 1  & $20$ & $0.43817 + 0.38594 \, i$ & $38$& $57$ & $85$ & $97$ & $119$  \\
 5 & $80$ & $0.88317 + 0.090624 \, i$& $ 56$ & $71$ & $96$ & $112$ & $136$\\
 10 & $100$ &  $0.89225 + 0.10828 \, i$ & $63$ & $75$ & $92$ & $103$ & $107$  \\
\hline
\end{tabular}	
\caption{Normalized surface conductivity $\zeta_0 \sigma$ that cancel the backscattering by a sphere of given $r/\lambda$, $\varepsilon_r=4$, and assuming an expansion order $N_{max}$. Corresponding  
forward-to-backward scattering ratios, calculated by using only $sf$ significant figures of $\sigma$.}
\label{tab:Back}
\end{table*}
\begin{figure*}[ht]
\centering
	\includegraphics[width=\textwidth]{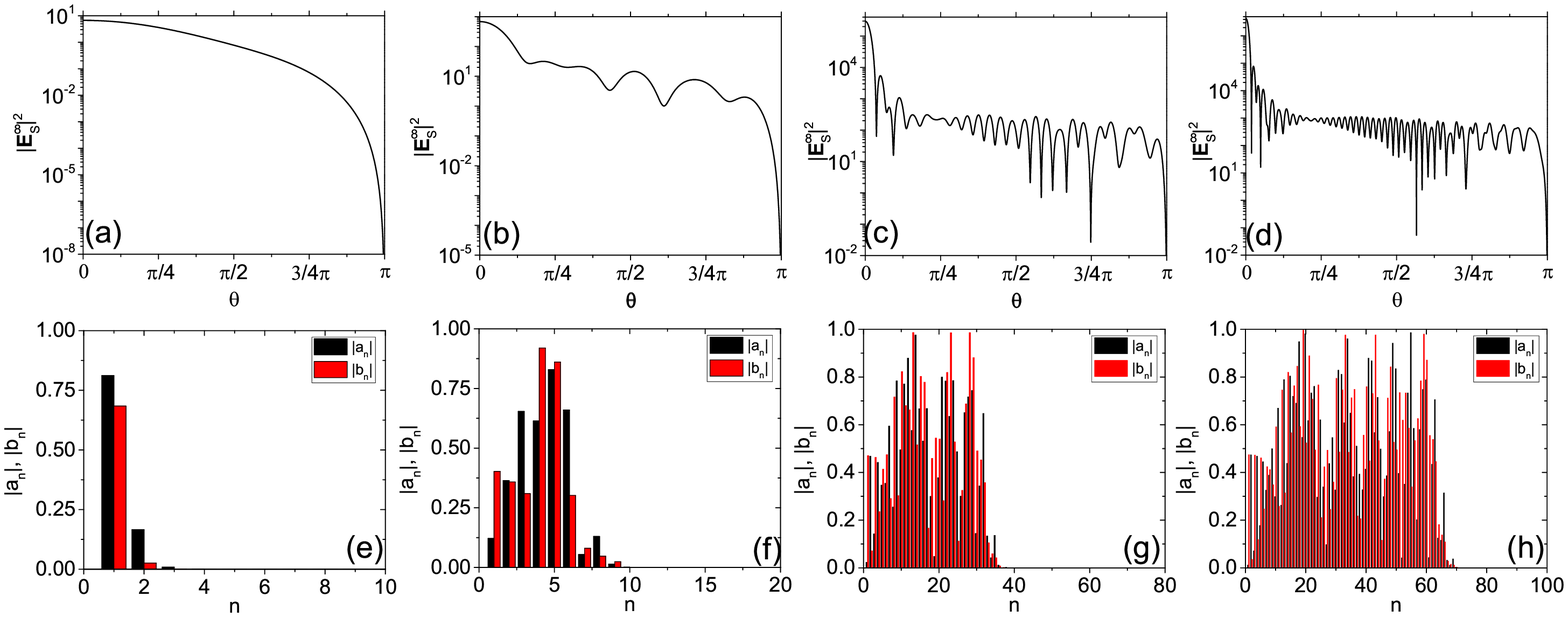}
\caption{Radiation diagram  as a function of the angle $\theta$ at $\phi = 0$ for a sphere with permittivity $\varepsilon_{r}=4$ and  $r/\lambda=1/4$ (a), $r/\lambda=1$ (b), $r/\lambda=5$ (c), $r/\lambda=10$ (d), coated by a surface of finite conductivity. The value of surface conductivity $\sigma$ is designed to enforce a vanishing backscattering. The corresponding magnitude of the Mie coefficients  (as defined by Eq. \ref{eq:MieCoefficients}) is shown in panels (e-h).}
\label{fig:Back}
\end{figure*}

By substituting the expansions \ref{eq:Cp},\ref{eq:Cm} into the homogeneous problem, we obtain, for a given indices pair $m$ and $n$, the equations:
\begin{equation}
 {\bf Q}^{\mathtt{TM}}
 \left[ \begin{array}{c}
A_{\substack{ e \\ o}mn} \\
D_{\substack{ e \\ o}mn} \end{array}	
  \right] = 
 \left[ \begin{array}{c}
  0 \\ 0 
 \end{array}\right] , \quad 
     {\bf Q}^{\mathtt{TE}} 
 \left[ \begin{array}{c}
B_{\substack{ e \\ o} mn}  \\
C_{\substack{ e \\ o} mn} \end{array}
  \right] =  \left[ \begin{array}{c}
  0 \\ 0 
 \end{array}\right], 
\label{eq:LinearProblem}
\end{equation}
\begin{equation}
   \begin{aligned}
      {\bf Q}^{\mathtt{TM}} & = \left[ \begin{array}{cc}
\h{x}{n} - i \left( \frac{\zeta_0 \sigma}{x} \right) \left[ x \, \h{x}{n} \right]' & -  \SqrtEps \J{ \SqrtEps x}{n} \\ 
\SqrtEps \left[ x \, \h{x}{n} \right]' &
 -\left[ \SqrtEps x \, \J{\SqrtEps x}{n} \right]'   \end{array}
  \right] \\
   {\bf Q}^{\mathtt{TE}} & =    \left[ \begin{array}{cc}
\h{x}{n}  & -\J{\SqrtEps x}{n} \\
 \left[ x \, \h{x}{n} \right]' + i \left( x \zeta_0 \sigma \right) \h{x}{n}  &
 -\left[ x \SqrtEps \, \J{x \SqrtEps	}{n} \right]'
  \end{array}
  \right]
 \end{aligned} 
 \label{eq:Mat}
\end{equation}
where the prime denotes differentiation with respect to the argument in parentheses, $j_n$ and $h_n^{\left(1 \right)} $ are the spherical Bessel and spherical Hankel functions, and $\zeta_0$ is the normalized characteristic impedance. Non trivial solutions of the linear problems described by Eqs. \ref{eq:LinearProblem}
are obtained by zeroing the determinant of the corresponding matrices shown in Eqs. \ref{eq:Mat}. 
The corresponding resonant surface conductivities  are
\begin{equation}
\begin{aligned}
 \sigma_n^{\texttt{TM}} &=- \frac{i x}{\zeta_0} \; \frac{h_n^{(1)}(x) \left[ \SqrtEps x j_n(\SqrtEps x) \right]' - \varepsilon_r j_n(\SqrtEps x) \left[ x h_n^{(1)}(x) \right]'}{\left[ \SqrtEps x j_n(\SqrtEps x) \right]' \left[ x h_n^{(1)}(x) \right]' }, \\
 \sigma_n^{\texttt{TE}} &= -\frac{i}{x \zeta_0} \frac{  h_n^{(1)}(x) \left[ \SqrtEps x j_n(\SqrtEps x) \right]'-j_n(\SqrtEps x) \left[ x h_n^{(1)}(x)\right]'}{h_n^{(1)}(x) j_n(\SqrtEps x)}.
\end{aligned}
\end{equation}
They only depend on $r/\lambda$ and $\varepsilon_r$. The eigenspaces corresponding to them are spanned by \begin{equation}
   {\bf E}_{S,pmn}^{\mathtt{TM}} = 
   \begin{cases}
     \Ni{3}{k_0 {\bf r}}{pmn} & \quad {\bf r} \in \Omega^- \\
 \frac{\SqrtEps \left[ x \, \h{x}{n} \right]'}{\left[ \SqrtEps x \, \J{\SqrtEps x}{n} \right]'} \Ni{1}{\SqrtEps  k_0 {\bf r}}{pmn} & \quad {\bf r} \in \Omega^+   
   \end{cases},
\end{equation}
\begin{equation}
   {\bf E}_{S,pmn}^{\mathtt{TE}} = 
   \begin{cases}
     \Mi{3}{k_0 {\bf r}}{pmn} & \quad {\bf r} \in \Omega^- \\
    \frac{\h{x}{n}}{\J{\SqrtEps x}{n}} \Mi{1	}{\SqrtEps  k_0 {\bf r}}{pmn} & \quad {\bf r} \in \Omega^+   
   \end{cases},
\end{equation}
where $m \in \mathbf{N}_0$ and $p \in \left\{ e, o \right\}$.

Starting from the electric field modes, we obtain the expansion of $\tilde{\bf E}_S$, solution of the non-homogeneous problem of Eqs. \ref{eq:MErot2}-\ref{eq:BC2}
\begin{equation}
   \tilde{\bf E}_S \left( {\bf r} \right) = 
\displaystyle\sum_{p m n } \left(    \frac{\sigma}{ \sigma_n^{\texttt{TM}} -  \sigma }  \, \mathcal{P}_{pmn}^{\mathtt{TM}} \, {\bf E}_{pmn}^{\mathtt{TM}}
+ \frac{ \sigma }{ \sigma_n^{\texttt{TE}} -    \sigma} \, \mathcal{P}_{pmn}^{\mathtt{TE}} \, {\bf E}_{pmn}^{\mathtt{TE}}
   \right)
   \label{eq:EsTilde}
\end{equation}
\begin{equation}
\mathcal{P}_{pmn}^{\mathtt{TM}} = \frac{\langle 
     \Etoto^* ,   {\bf E}_{pmn}^{\mathtt{TM}} \rangle_\Sigma}{\langle   \left( {\bf E}_{pmn}^{\mathtt{TM}} \right)^*,   {\bf E}_{pmn}^{\mathtt{TM}} \rangle_\Sigma} , \;
\mathcal{P}_{pmn}^{\mathtt{TE}} = \frac{\langle  
  \Etoto^*,  {\bf E}_{pmn}^{\mathtt{TE}} \rangle_\Sigma}{\langle   \left( {\bf E}_{pmn}^{\mathtt{TE}}\right)^*,  {\bf E}_{pmn}^{\mathtt{TE}} \rangle_\Sigma} .
\end{equation}
where 
$
 \langle {\bf A}, {\bf B} \rangle_\Sigma = \int_\Sigma \left( \overleftrightarrow{\mbox{T}} {\bf A} \right)^* \cdot  \left( \overleftrightarrow{\mbox{T}} {\bf B} \right)\, dS.
$
In the case of a $x$-polarized  plane-wave excitation,  propagating along the $z$-axis,  only the coefficients $\mathcal{P}_{e1n}^{\mathtt{TM}}$ and $\mathcal{P}_{o1n}^{\mathtt{TE}}$ are non-vanishing and have the closed-form expression:
\begin{equation}
  \mathcal{P}^{\mathtt{TM}}_{e1n} = i E_n \left( a_n^0 + \frac{\left[ x \, \J{x}{n} \right]'}{\left[ x \, \h{x}{n} \right]'} \right),
  \mathcal{P}^{\mathtt{TE}}_{o1n} = - E_n \left( b_n^0 - \frac{\J{x}{n}}{\h{x}{n}}\right).
\end{equation}
The scattered field in $ \Omega^-$ is  obtained by adding Eqs. \ref{eq:PW_VSWF} and \ref{eq:EsTilde}:
\begin{equation}
   {\bf E}_S^- \left( {\bf r} \right)  =  \displaystyle\sum_{n=1}^\infty  E_n \left(  i a_n \Ni{3}{k_0 {\bf r}}{e1n} -b_n \Mi{3}{k_0 {\bf r}}{o1n}   \right),
  \label{eq:ScattExp}
\end{equation}
\begin{equation}
\begin{aligned}
  a_n &= a_n^0 + \frac{   \sigma }{  \sigma_n^{\mathtt{TM}} -   \sigma  }\left(  a_n^0 +	 \frac{\left[ x \, \J{x}{n} \right]'}{\left[ x \, \h{x}{n} \right]'}\right)	
   \\
b_n &= b_n^0 + \frac{   \sigma }{ \sigma_n^{\mathtt{TE}} -   \sigma }\left( b_n^0 - \frac{ \J{x}{n} }{ \h{x}{n}}\right)
\end{aligned}
\label{eq:MieCoefficients}
\end{equation}
In Eq. \ref{eq:ScattExp}, the material of the coating only appears through $\sigma$ in the factors $\frac{   \sigma  }{ \sigma_n^{\mathtt{TM}} -   \sigma }$ and
 $\frac{  \sigma  }{ \sigma_n^{\mathtt{TE}} -   \sigma }$. Therefore, Eq. \ref{eq:ScattExp}  separates the role of geometry and size from the role played by the material. We note that the expansion \ref{eq:ScattExp} is analytical.
 
\begin{table*}[ht]	
\centering
\begin{tabular}{ccccccc} 
\hline
 \multirow{2}{*}{$r/\lambda$} & \multirow{2}{*}{$N_{max}$} &  \multirow{2}{*}{$\zeta_0 \sigma$} & 
   \multicolumn{4}{c}{$ \left\| \mathbf{E} \left( \theta = 0 \right)/\mathbf{E}\left( \theta = \pi \right) \right\| (dB) $}    \\
 \cmidrule(l){4-7}
    & & & $sf=2$&$sf=3$ &    $sf=4$& $sf=5$ \\
   \hline
 $0.25$ & $10$  & $-1.2536 + 1.0063 \, i$ & $-27$ & $-48$ & $-66$ & $-80$ \\
 $1$ & $20$ & $-2.0095 + 0.82776 \, i$ & $-34$ & $-59$ & $-59$ & $-84$  \\
 $5$ & $80$ & $-2.0482 + 13.704 \, i$ & $-28$  & $-28$ & $-28$ & $-53$  \\
\hline
\end{tabular}	
\caption{Normalized surface conductivity $\zeta_0 \sigma$ that cancel the forward scattering by a sphere of given $r/\lambda$, with $\varepsilon_r=4$ and assuming an expansion order $N_{max}$. Corresponding  
forward-to-backward scattering ratios, calculated by using only $sf$ significant figures of $\sigma$.}
\label{tab:Fdw}
\end{table*}
\begin{figure*}[ht]
\centering
	\includegraphics[width=0.75\textwidth]{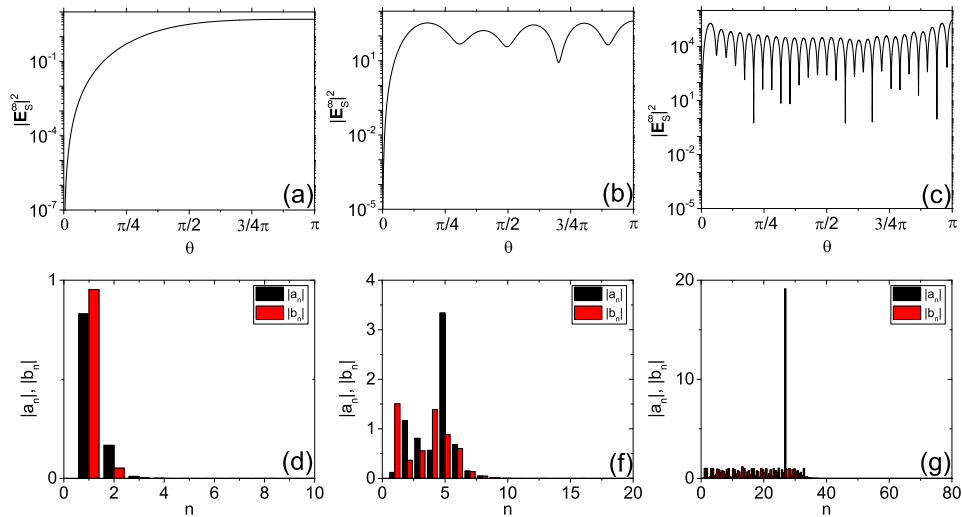}
\caption{Radiation diagram as a function of the angle $\theta$ at $\phi = 0$ for a sphere with permittivity $\varepsilon_{r}=4$ and  $r/\lambda=1/4$ (a), $r/\lambda=1$ (b), $r/\lambda=5$ (c), coated by a surface of finite conductivity. The value of surface conductivity $\sigma$ is designed to enforce a vanishing forward-scattering. The corresponding magnitude of the Mie coefficients  (as defined by Eq. \ref{eq:MieCoefficients}) is shown in panels (d-g).}
\label{fig:Fdw}
\end{figure*}

We now show how to cancel either the back or forward scattering of a homogeneous sphere by designing the permittivity of its surface coating. The radiation pattern is defined as
\begin{equation}
 \mathbf{E}_S^\infty \left( \theta, \phi \right) = \displaystyle\lim_{r \rightarrow \infty}  \left[ r{e^{- i k_0 r}} {\bf E}_S^- \right],
\end{equation} 
where $\theta$ and $\phi$ are the polar and azimuthal angles, respectively.
Due to symmetry considerations, the only non-vanishing component of the radiation pattern at both the back ($\theta = \pi$) and the forward ($\theta = 0$) directions  is the $\theta$ component $\mathbf{E}_S^\infty \cdot {\bf i}_\theta$. Therefore, we have to find the zeros of $\mathbf{E}_S^\infty \cdot {\bf i}_\theta$ as a function of $\sigma$, where $\mathbf{E}_S^\infty \cdot {\bf i}_\theta$ is expressed as:

\begin{equation}
\label{eq:diffSca}
   {\bf E}_S^\infty \cdot {\bf i}_\theta =\sum_{n=1}^{\infty} E_n \left[ i\,a_n \Ni{\infty}{\theta,\phi}{e1n}  \cdot {\bf i}_\theta -b_n 
\Mi{\infty}{\theta,\phi}{o1n} \cdot {\bf i}_\theta   
   \right],
\end{equation}
where $a_{n}$ and $b_{n}$ are defined in Eqs. \ref{eq:MieCoefficients}, and
$ \Mi{\infty}{\theta,\phi}{o1n} = \displaystyle\lim_{r \rightarrow \infty}  \left[ k_0 {r} {e^{- i k_0 r}} \Mi{3}{\theta,\phi}{o1n} \right]$, $ \Ni{\infty}{\theta,\phi}{e1n} = \displaystyle\lim_{r \rightarrow \infty}  \left[ k_0 {r} {e^{- i k_0 r}} \Ni{3}{\theta,\phi}{e1n} \right].$
We substitute Eqs. \ref{eq:MieCoefficients} into Eq. \ref{eq:diffSca}, then we fix the values of $r/\lambda$, $\varepsilon_r$, $\phi=0$, and $\theta=\pi$ for the backscattering cancellation,  and $\theta= 0$ for the forward scattering cancellation. We then truncate the sum to $N_{max}$. The resulting expression of ${\bf E}_S^\infty \cdot {\bf i}_\theta $ is a complex-valued function of the complex variable $\sigma$. We put all the terms of the resulting sum over a common denominator, obtaining, in this way, a rational function of $\sigma$. Eventually, we zero  the resulting numerator, which is a polynomial in $\sigma$ of order $2 N_{max}$. Several solutions can be found.

In Tabs. \ref{tab:Back} and \ref{tab:Fdw} we show the values of $\sigma$ with five significant figures, that guarantee the back and the forward scattering cancellation, for different values of $r/\lambda$ and for $\varepsilon_r = 4$. In addition, for each scenario, we show the achieved values of forward-to-backward scattering ratios, i.e.
\begin{equation}
   \rho =  {\left\| {\bf E}_S^\infty \left( \theta=\pi ,\phi=0 \right) \right\|}/{ \left\| {\bf E}_S^\infty \left( \theta=0 ,\phi=0 \right) \right\|},
\end{equation}
as a function of the number of significant figures $sf$ of $\sigma$. Although $\rho$ is highly sensitive to $sf$, we note that, even with $sf=2$, we achieve a good directional scattering cancellation. All the values of $\sigma$ that guarantee forward scattering cancellation have negative real part, which correspond to active surfaces, consistently with the optical theorem \cite{bohren08}.

For each of the scenarios of Tabs. \ref{tab:Back}-\ref{tab:Fdw}, we plot in Fig. \ref{fig:Back} (a-d) and \ref{fig:Fdw} (a-c) the corresponding squared magnitude of the radiation pattern as a function of $\theta$ for $\phi=0$, while, in Fig. \ref{fig:Back} (e-h) and \ref{fig:Fdw} (d-g), we plot the magnitude of the Mie coefficients. We note that the scattering cancellation is not a result of destructive interference between magnetic and electric multipoles of same order but is a global interference among all the scattering orders.

The cancellation of the forward scattering may be very challenging. Since the surface that guarantees forward scattering cancellation is necessarily active, $\sigma$ lies either on the II or on the III quadrant of the complex plane. Therefore, the zeros may be positioned very close to one of the  eigen-conductivities. This fact has two important consequences. First, finding the roots of the polynomial is numerically harder, and an increased working precision may be needed. Second, the forward scattering may be very sensitive to perturbation of $\sigma$ in the neighbourhood of the solution. This is exactly what happens for $r/\lambda=5$ where the value of $\sigma$, shown in Tab \ref{tab:Fdw}, is very close to the eigen-conductivity $\zeta_0 \sigma_{27}^\mathtt{TM}  = -1.9976 + 13.614 \, i$, and the electric multipole $n=27$ is dominant, as shown in Fig. \ref{fig:Fdw} (g).


\begin{thebibliography}{10}
\newcommand{\enquote}[1]{``#1''}

\bibitem{kerker1983electromagnetic}
M.~Kerker, D.-S. Wang, and C.~Giles, JOSA \textbf{73}, 765 (1983).

\bibitem{gomez2011electric}
R.~Gomez-Medina, B.~Garcia-Camara, I.~Su{\'a}rez-Lacalle, F.~Gonz{\'a}lez,
  F.~Moreno, M.~Nieto-Vesperinas, and J.~J. S{\'a}enz, Journal of Nanophotonics
  \textbf{5}, 053512 (2011).

\bibitem{nieto2011angle}
M.~Nieto-Vesperinas, R.~Gomez-Medina, and J.~Saenz, JOSA A \textbf{28}, 54
  (2011).

\bibitem{geffrin2012magnetic}
J.-M. Geffrin, B.~Garc{\'\i}a-C{\'a}mara, R.~G{\'o}mez-Medina, P.~Albella,
  L.~Froufe-P{\'e}rez, C.~Eyraud, A.~Litman, R.~Vaillon, F.~Gonz{\'a}lez,
  M.~Nieto-Vesperinas \emph{et~al.}, Nature communications \textbf{3}, 1171
  (2012).

\bibitem{fu2013directional}
Y.~H. Fu, A.~I. Kuznetsov, A.~E. Miroshnichenko, Y.~F. Yu, and
  B.~Luk’yanchuk, Nature communications \textbf{4}, 1527 (2013).

\bibitem{Person13}
S.~Person, M.~Jain, Z.~Lapin, J.~J. Saenz, G.~Wicks, and L.~Novotny, Nano
  letters \textbf{13}, 1806 (2013).

\bibitem{liu2018generalized}
W.~Liu and Y.~S. Kivshar, Optics express \textbf{26}, 13085 (2018).

\bibitem{Forestiere16}
C.~Forestiere and G.~Miano, Phys. Rev. B \textbf{94}, 201406 (2016).

\bibitem{pascale2017spectral}
M.~Pascale, G.~Miano, and C.~Forestiere, JOSA B \textbf{34}, 1524 (2017).

\bibitem{xia2014two}
F.~Xia, H.~Wang, D.~Xiao, M.~Dubey, and A.~Ramasubramaniam, Nature Photonics
  \textbf{8}, 899 (2014).

\bibitem{bohren08}
C.~F. Bohren and D.~R. Huffman, \emph{Absorption and scattering of light by
  small particles} (John Wiley \& Sons, 2008).

\end{thebibliography}
\end{document}